\documentclass[doublecol]{epl2} 

\usepackage{graphicx}
\usepackage{amsmath}
\usepackage{amssymb}
\usepackage{bm}

\newcommand{\be}{\begin{equation}}
\newcommand{\ee}{\end{equation}}
\newcommand{\bea}{\begin{eqnarray}}
\newcommand{\eea}{\end{eqnarray}}
\newcommand{\bei}{\begin{itemize}}
\newcommand{\eei}{\end{itemize}}

\newcommand{\openone}{1}

\newcommand{\Er}{E_{\rm R}}

\title{Optical Flux Lattices for Two-Photon Dressed States}

\author{Nigel R. Cooper\inst{1} \and Jean Dalibard\inst{2}}
\shortauthor{N.R. Cooper and J. Dalibard}

\institute{ \inst{1}T.C.M. Group, Cavendish Laboratory, J.~J.~Thomson Avenue, Cambridge CB3~0HE, United Kingdom\\
\inst{2} Laboratoire Kastler Brossel, CNRS, UPMC, Ecole Normale Sup\'erieure, 24 rue Lhomond, 75005 Paris, France
}

\abstract{We describe a simple scheme by which ``optical flux lattices'' can be implemented in ultracold atomic gases using two-photon dressed states.  This scheme can be applied, for example, to the ground state hyperfine levels of commonly used atomic species.  The resulting flux lattices simulate a magnetic field with high mean  flux density, and have low energy bands analogous to the lowest Landau level.  We show that in practical cases the atomic motion significantly deviates from the adiabatic following of one dressed state, and that this can lead to significant interactions even
for fermions occupying a single band.  Our scheme allows experiments on cold atomic gases to explore strong correlation phenomena related to the fractional quantum Hall effect,
  both for fermions and bosons.}

\begin{document}

\maketitle

There has long been an endeavour in the field of ultracold atomic
gases to achieve an experimental regime in which strong correlation
phenomena associated with the fractional quantum Hall (FQH) effect
can appear\cite{Bloch:2008,Fetter:2009,Cooper:2008}. Central to this goal is the
formation of Landau levels: the highly degenerate energy levels of a
two dimensional charged particle in a uniform magnetic field.  Owing
to this degeneracy, particles occupying a single Landau level are
susceptible to interparticle interactions. These can give rise to
strongly correlated FQH states when the particle density is comparable
to the density of magnetic flux quanta $n_\phi$. Existing experimental
studies of FQH states in semiconductors involve fermions
(electrons). However it is expected that related strong-correlation
phenomena can arise also for bosonic species of cold
atoms\cite{Cooper:2008}.

The formation of Landau levels for neutral atoms requires the creation
of an effective magnetic field.  To date, this has been achieved in
experiments on ultracold atoms either by using
rotation\cite{Fetter:2009,Cooper:2008}, or by using optical dressing
to generate effective gauge fields\cite{Lin2009b}.  However, in both
cases the magnetic flux densities $n_\phi$ achieved so far are too low
to bring a large atom cloud into the fractional quantum Hall regime.
Theoretical proposals for methods to generate effective magnetic
fields for atoms on deep optical lattices hold promise to achieve high
flux densities, on the order of the inverse optical wavelength squared
$n_\phi\sim 1/\lambda^2$ (for reviews of these proposals see, {\it
  e.g.}  \cite{Lewenstein:2007,Dalibard:2010}). At these flux
densities, interaction energy scales would be sufficiently large to
allow experimental studies even in the FQH regime.

A recent theoretical proposal has shown that effective magnetic fields
with high flux density $n_\phi\sim 1/\lambda^2$ can also be generated
for optically dressed states without the use of deep optical lattices. 
In this so-called ``optical flux lattice'' scheme\cite{Cooper:2011} a periodic
effective magnetic field with non-zero average is generated by
arranging that the spatial dependence of the Rabi coupling and a
state-dependent potential are in register with each other.  The
implementation proposed in Ref.\cite{Cooper:2011} involves a
single-photon coupling, suitable for Ytterbium or alkaline earth
atoms, and requires locking the positions of standing waves of a laser
at the coupling frequency with those of a laser at an ``anti-magic''
frequency.

In this paper, we show how optical flux lattices can be generated by
using two-photon dressed states.  Our scheme also requires two
laser frequencies, but it is more robust than that of
\cite{Cooper:2011}: the Rabi coupling and state-dependent potential
are automatically in register.  Importantly, our scheme can be applied
to commonly used atomic species, including alkali atoms in hyperfine
levels of any angular momentum $F$.  We describe the bandstructures
for representative cases $F=1/2$ and $F=1$.  We show that there appear
low energy bands that are analogous to the lowest Landau level: of
narrow width in energy and with nonzero Chern
number\cite{Thouless:1982a}.  Owing to the narrow energy width,
particles occupying these bands are susceptible to interactions and to
the formation of strongly correlated FQH states.  We show that, even
for fermions interacting with contact interactions, there remain
significant interparticle interactions within this low energy band.
Thus, our scheme will allow experiments on cold atomic gases to
explore strong correlation phenomena related to the fractional quantum
Hall effect for both fermions and bosons.

In the first part of this paper we consider an atomic species with a ground level $g$ of angular momentum $J_g=1/2$. Examples of atoms in this category that have already been laser-cooled  are  $^{171}$Yb or $^{199}$Hg (level $6\,^1$S$_0$)  \cite{Taie:2010,Yi:2011}. The atoms are irradiated by laser waves of frequency $\omega_L$ that connect $g$ to an excited state $e$ also with angular momentum $J_e=1/2$. For  Ytterbium and Mercury atoms, we can choose $e$ to be the first excited level $6\,^3$P$_0$ entering in the so-called `optical clock' transition. The very long lifetime of $e$  ($\sim 10$~s for Yb \cite{Porsev:2004} and $\sim 1$~s for Hg \cite{Bigeon:1967}) guarantees that heating due to random spontaneous emissions of photons is negligible on the time scale of an experiment. Another possible choice could be $^6$Li atoms, but we estimated in this case a photon scattering rate that is too large to maintain the gas at the required low temperature.

We assume that the atomic motion is restricted to the $xy$ plane and described by the Hamiltonian
\be 
\hat{H} = \frac{{{\bm p}}^2}{2M}
\hat \openone + \hat{V}({\bm r}) \ ,
\label{eq:ham} 
\ee 
where $M$ is the atomic mass and ${\bm p}$ its momentum. The matrix $\hat V$ acts in the Hilbert space describing the internal atomic dynamics. For an off-resonant excitation, we can assume that the population of $e$ is negligible at all times, so that $\hat V$ is a $2\times2$ matrix acting the $g_\pm$ manifold \cite{cohe92}. Its coefficients depend on the local value of the laser electric field, which we characterize by the Rabi frequencies $\kappa_m$, $m=0,\pm 1$, where $m\hbar$ is the angular momentum along $z$ gained by the atom when it absorbs a photon.

\begin{figure}[t]
\begin{center}
\includegraphics[width=88mm]{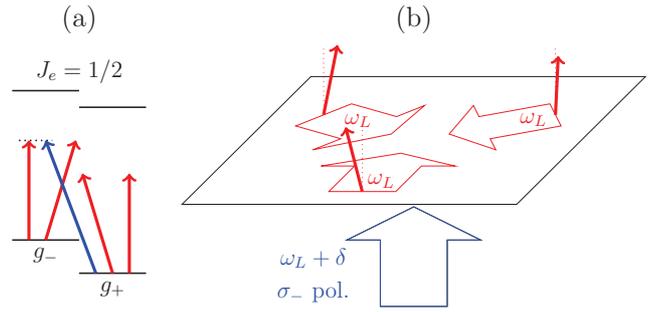}
\caption{(Colour on line) (a) A ground level with angular momentum $J_g=1/2$ is coupled to an excited level also with angular momentum $J_{e}=1/2$ by laser beams at frequency $\omega_L$ and $\omega_L+\delta$. The Zeeman splitting between the two ground states $g_\pm$ is $\delta$. (b) Three linearly polarized beams at frequency $\omega_L$ with equal intensity and with wave vectors at an angle of $2\pi/3$ propagate in the $xy$ plane. The fourth, circularly polarized beam  at frequency  $\omega_L+\delta$ propagates along the $z$ axis.}
\label{fig:scheme}
\end{center}
\end{figure}

In order to increase our control on the spatial variations of $\hat V$, we suppose that a magnetic field parallel to the $z$-axis lifts the degeneracy between the states $g_\pm$.  The resulting splitting $\delta$ is supposed to be much larger than the $\kappa_m$'s. Hence for a monochromatic laser excitation at frequency $\omega_L$, the off-diagonal matrix elements $V_{+-}$ and $V_{-+}$ are negligible compared to the diagonal ones. However we also assume that another laser field at frequency $\omega_L+\delta$, propagating along the $z$ axis with $\sigma_-$ polarization (\emph{i.e.} $m=-1$ with the notation above), is shone on the atoms. The association of this field with the $\pi$ component ($m=0$) of the light at $\omega_L$ provides the desired resonant Raman coupling between $|g_\pm\rangle$ (figure \ref{fig:scheme}a). Using standard angular momentum algebra we find in the $\{|g_+\rangle, |g_-\rangle\}$ basis:
\begin{equation}
\hat V=\frac{\hbar \kappa_{\rm tot}^2}{3\Delta}\hat \openone+
\frac{\hbar}{3\Delta }\begin{pmatrix}
 |\kappa_-|^2-|\kappa_+|^2 & E \kappa_0\\
 E \kappa_0^* &  |\kappa_+|^2-|\kappa_-|^2
\end{pmatrix}.
\label{eq:matrixV}
\end{equation}
Here $\kappa_{\rm tot}^2=\sum_m |\kappa_m|^2$, $\Delta=\omega_L-\omega_A$, where  $\omega_A$ is the atomic resonance frequency, and we assume $|\Delta|\gg |\delta|,|\kappa_m|$. The quantity $E$ characterizes the field of the additional laser at $\omega_L+\delta$. This beam is assumed to be a plane wave propagating along $z$, so that $E$ is a uniform, adjustable coupling. The a.c. Starkshift due to this additional laser is incorporated in the definition of $\delta$.

We consider the laser configuration represented in Fig.\,\ref{fig:scheme}(b). The laser field at frequency $\omega_L$ is formed by the superposition of three plane travelling waves of equal intensity with wavevectors ${\bm k}_i$ in the $xy$ plane. We focus on a situation of triangular symmetry, in which the three beams  make an angle of $2\pi/3$ with each other,  ${\bm k}_1 =- k/2\,\left(\sqrt{3},1,0\right)$, ${\bm k}_2 = k/2\,\left(\sqrt{3},-1,0\right)$ and ${\bm k}_3 = k\left(0,1,0\right)$. Each beam is linearly polarized at an angle $\theta$ to the $z$-axis, which leads to
\be
\label{eq:kappa}
{\bm \kappa}  = \kappa \sum_{i=1}^3 e^{i{\bm k}_i\cdot {\bm r}} \left[ \cos\theta\, \hat{{\bm z}}
+ \sin\theta\, (\hat{{\bm z}}\times \hat{{\bm k}}_i)\right],
\ee
where $\kappa$ is the Rabi frequency of a single beam. In the following we denote ${\cal V}=\hbar \kappa^2/(3\Delta)$ the energy associated with the atom-light interaction  and $\epsilon=E/\kappa$ the relative amplitude  of the $\omega_L+\delta$ field with respect to the $\omega_L$ field. The recoil energy $\Er=\hbar^2k^2/2M$ sets the characteristic energy scale of the problem. 

The coupling $\hat V$ is written in Eq.\,(\ref{eq:matrixV}) as the sum of the scalar part $\hbar \kappa_{\rm tot}^2/(3\Delta)\,\hat \openone$ and a zero-trace component that can be cast in the form $\hat W=\hat {\bm \sigma}\cdot \bm B/2$, 
where the $\hat \sigma_i$ are the Pauli matrices ($i=x,y,z$). For $E\neq 0$ and $\sin 2\theta \neq 0$, the coupling ${\bm B}$ is everywhere non-zero.  Suppose that the atom is prepared in the local eigenstate $|\chi (\bm r)\rangle$ of $\hat W$, with a maximal angular momentum projection along ${\bm n} = -{\bm B}/|{\bm B}|$. Suppose also that it moves sufficiently slowly to follow adiabatically this eigenstate, which is valid when ${\cal V}\gg \Er $. This leads to the Berry's-phase-related gauge potential $i\hbar \langle \chi |\bm \nabla \chi\rangle$,  representing a non-zero effective magnetic flux density \cite{Berry:1984}. For most optical lattice configurations, the periodic variation of the atom-laser coupling leads to a zero net flux of the effective field through a unit cell of the lattice. Indeed the periodicity of the eigenstate $|\chi\rangle$ entails that the line integral of $\langle \chi |\bm \nabla \chi\rangle$ on the edges of the unit cell vanishes. However this reasoning may fail if the vector potential has singularities inside the cell:  this is the so-called \emph{flux lattice} scheme \cite{Cooper:2011}.

The singularities of $\langle \chi |\bm \nabla \chi\rangle$ occur at
places where the $g_+$--$g_-$ coupling [$\propto \kappa_0$, see
Eq.\,(\ref{eq:matrixV})] vanishes. Here, the zeroes of $\kappa_0$ are located
on a honeycomb lattice, with a distance of $2\lambda/(3\sqrt 3)$
between nearest neighbours, where $\lambda=2\pi/k$ \cite{Lee:2009}. The
unit cell of this hexagonal lattice has an area $A_{\rm
  cell}=2\lambda^2/(3\sqrt 3)$, and contains two inequivalent sites
that correspond to vortices and antivortices of $(n_x,n_y)$,
respectively. We will introduce later a gauge transform that allows
one to use the same unit cell for the periodic coupling $\hat V$. At
the location of the (anti)vortices of $(n_x,n_y)$, the modulus of
$B_z\propto |\kappa_+|^2-|\kappa_-|^2$ is maximal and its sign
alternates, leading to $n_z = \pm 1$. Because the sign of $n_z$
differs for vortices and antivortices, each of these regions wraps the
Bloch sphere in the same sense providing the desired rectification of
the magnetic flux.  The net result is that ${\bm n}$ wraps the Bloch
sphere once within the unit cell, corresponding to $N_\phi = 1$ flux
quanta. This net flux $N_\phi=1$ is a robust feature of the optical coupling, not relying on fine-tuning of parameters or on the control of the relative phase of the laser beams. Reversal of the magnetic flux ($N_\phi=-1$) can be achieved by changing the sign of the detuning $\Delta$, or replacing the $\sigma_-$ laser at frequency $\omega_L+\delta$ by a $\sigma_+$ laser at frequency $\omega_L-\delta$.

Our laser configuration is reminiscent of the scheme studied in
\cite{Dudarev:2004}. However in that work only the three beams at
frequency $\omega_L$ and propagating in the $xy$ plane were
considered. The off-diagonal coupling in $\hat V$ is in this case
$V_{+-}\propto \kappa_+\kappa_0^*+\kappa_0\kappa_-^*$, and one can
check the desired rectification of the magnetic flux at the zeroes of
$V_{+-}$ does not occur. Therefore, although the
Berry's-phase-related magnetic field found in \cite{Dudarev:2004} is
locally non zero, the flux per unit cell vanishes.

In the adiabatic limit, valid for ${\cal V}\gg \Er $
\cite{Dalibard:2010}, the atoms are strongly confined to the sites
where $\hbar \kappa_{\rm tot}^2/(3\Delta)-|\bm B|/2$ is
minimum. Depending on the sign of $\Delta$ and the values of $\theta$
and $E$, these sites coincide either with the vertices of the hexagons
of the honeycomb lattice introduced above, or with the centre of these
hexagons, forming thus a triangular lattice. Both of these
tight-binding limits are situations in which the number of flux quanta
per hexagonal (resp. triangular) plaquette is integer
(resp. half-integer), \emph{i.e.} time-reversal symmetry is
preserved\footnote{In both cases, the unit cell of the lattice contains
$N_\phi=1$ flux quantum.}.

In particular in this adiabatic limit the various energy bands cannot
have a non-zero Chern number \cite{Thouless:1982a}. However, an
important aspect of the physics of optical flux lattices is that they
do not rely on the adiabatic limit, and can be applied in regimes of
intermediate coupling ${\cal V}\sim \Er $.  Indeed the optical flux
lattices described here, and those of \cite{Cooper:2011}, lead to
bands with non-zero Chern number in the weak coupling limit
${\cal V}\lesssim \Er $. In this regime, a tight-binding model is
inappropriate; departures from this (with next nearest neighbour
tunnelling and beyond) allow the atoms to perform loops that enclose
flux that breaks time-reversal symmetry.

We have calculated the band structures for arbitrary ${\cal V}/ \Er $ by numerical diagonalization of the Hamiltonian (\ref{eq:ham}).  To do so, it is helpful to expose the full translational symmetry of the
system. The matrix $\hat V$ is invariant under translations by the vectors $\lambda/\sqrt 3\,(\pm 1,\sqrt 3,0)$, corresponding to a `naive' unit cell of area $2\lambda^2/\sqrt 3$. One can reduce the size of the unit cell by taking advantage of a unitary transformation by $\hat{U} \equiv \exp(-i
  {\bm k}_3 \cdot {\bm r}\, \hat{\sigma}_z/2)$.
 This state-dependent gauge
transformation gives a transformed Hamiltonian
\begin{eqnarray}
\label{eq:gauge}
\hat{H}' & = & \hat{U}^\dag \hat{H} \hat{U} 
  =  
 \frac{(\hat{{\bm p}}- \hat{\sigma}_z \hbar {\bm k}_3/2)^2}{2M}
+   \hat{V}' ,
\end{eqnarray}
where $\hat{V}' =  \hat{U}^\dag \hat{V} \hat{U}$ is of the same form as $\hat{V}$, but with $\kappa_0$ replaced by $ \kappa_0' =  e^{-i{\bm k}_3\cdot {\bm r}} \kappa_0$.
The advantage is that the coupling $\hat{V}'$ involves only the
momentum transfers ${\bm K}_{1,2} \equiv {\bm k}_{1,2} - {\bm
  k}_3$. Using the momentum transfers to define the reciprocal lattice
vectors leads to the largest possible Brillouin zone, and smallest
real space unit cell, which coincides with the unit cell of the scalar potential described above, with area $A_{\rm cell}=2\lambda^2/(3\sqrt 3)$.

\begin{figure}[t]
\includegraphics[width=0.9\columnwidth]{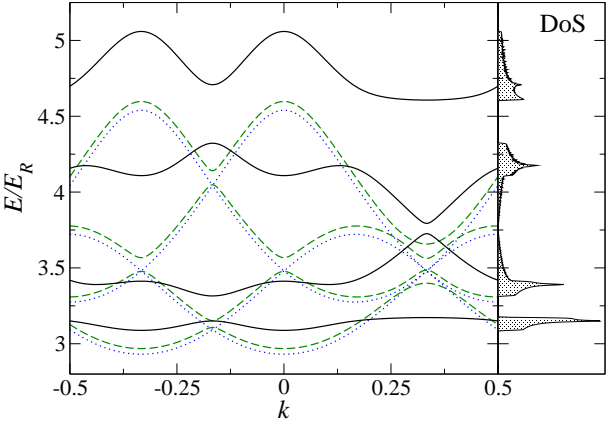}
\caption{(Colour online) Left-hand panel: Bandstructure for $F=1/2$ for ${\cal V}=1.8\,\Er $ along a path ${\bm k} = k({\bm K}_2-2{\bm K}_1) - {\bm k}_3/2$ through the Brillouin zone.   For $\epsilon =\theta =0$ (dotted blue line), the decoupled $m= \pm 1/2$ states each have two Dirac points. For weak coupling 
$\epsilon=\theta=0.1$ (dashed green line) the Dirac points split in a manner that breaks time-reversal symmetry, giving the lower two bands a net Chern  number of 1. For intermediate coupling $\epsilon=0.4$, $\theta=0.3$ (solid black line) the lowest energy band has Chern number 1. Right-hand panel: The density of states for  ${\cal V}=1.8\,\Er $, $\epsilon=0.4$ and $\theta=0.3$. 
\label{fig:dirac}}
\end{figure}

For the special case $\epsilon = \theta = 0$, the atoms only
experience the scalar potential $\propto|\kappa_0|^2$. For $\Delta
>0$, this situation was investigated in detail in \cite{Lee:2009}, in
connection with graphene physics. As explained above, $|\kappa_0|^2$
has minima at the sites of a honeycomb lattice. For ${\cal V} \gtrsim
\Er $ one therefore expects the atoms in both $g_\pm$ states to have
the low-energy bands of the honeycomb lattice with two Dirac points in
the Brillouin zone.  One difference with respect to graphene is that,
due to the state-dependent gauge transformation $\hat{U}$, the momenta
of the two spin states are offset by $\pm {\bm k}_3/2$. This has the
effect that there are three special points in the Brillouin zone: a
Dirac point for $g_+$, a Dirac point for $g_-$, and a (coincident)
Dirac point for both $g_\pm$. The bands are shown in
Fig.\,\ref{fig:dirac} with dotted lines for a cut through the Brillouin
zone that passes through these three points.

\begin{figure}[t]
\includegraphics[width=0.98\columnwidth]{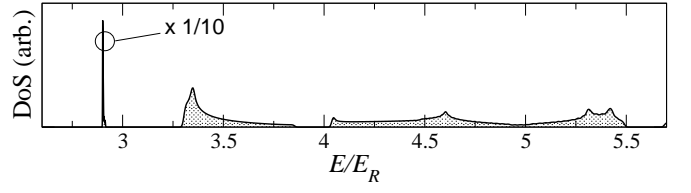}
\caption{Density of states for  $F=1/2$ for ${\cal V}=2\Er $,
$\theta = \pi/4$, $\epsilon=1.3$. The lowest band has Chern number 1, a width of  about
$0.01E_R$, and is well separated in energy from the next band. 
The density of states for the lowest band has been rescaled by $1/10$.
\label{fig:sharp}}
\end{figure}

Perturbations to this decoupled limit cause the Dirac points to split,
with consequences that depend on the symmetries that are broken and
the relative sizes of the perturbations\cite{Haldane:1988}.  For small
non-zero $\theta$, the terms $|\kappa_+|^2-|\kappa_-|^2\propto
\theta^2$ in (\ref{eq:matrixV}) break the sublattice symmetry of the
honeycomb lattice for both $g_\pm$. 
Gaps
then open in the spectrum at the locations of the Dirac points. For
$\epsilon=0$ the lower two bands separate from the upper two bands in
such a way that both pairs of bands are topologically trivial, that is
each pair has net Chern number of zero. When both $\epsilon$ and
$\theta$ are non-zero, the optical dressing leads to a net flux
through the unit cell, indicating time-reversal symmetry breaking.
Indeed, we find that for $\epsilon,\theta \neq 0$ the bands can
acquire non-zero Chern numbers.  Specifically, in the perturbative
limit ($\epsilon,\theta \ll 1$) the Dirac points split in such a way
that the lower two bands have a net Chern number of 1 provided
$\theta^2/\epsilon$ is sufficiently small. When $\theta^2/\epsilon$
exceeds a certain value ($\simeq 0.19$ for ${\cal V} = 1.8\,\Er$)
there is a transition to the topologically trivial case described
above.  Beyond the perturbative limit, as the couplings $\epsilon$ and
$\theta$ increase, the splitting between the lower two bands increases
and the lowest energy band evolves into a narrow band with Chern
number 1.  An example is shown by the solid lines and density of
states in Fig.\,\ref{fig:dirac}, for which the lowest band has a width
$\Delta E\simeq 0.1\,\Er $.  The optical dressing (\ref{eq:matrixV})
affords a great deal of freedom to tune parameters to reduce the ratio
of the bandwidth of the lowest band to the gap to the next band. For
example, for ${\cal V}=2\,\Er $, $\theta = \pi/4$, $\epsilon=1.3$,
the lowest band has a width of only $\Delta E \simeq 0.01E_R$ and is
separated from the next band by about $0.4\,\Er$ (See
Fig.\,\ref{fig:sharp}).

It is important to emphasize that the formation of this narrow low
energy band is not simply due to compression into a tight-binding
band\footnote{For a tight binding band in the limit of vanishing
  tunnel coupling when the Wannier states become exponentially
  localized, the Chern number of the band must be
  zero\cite{Thouless:1984}, or a set of bands with net Chern number
  zero must become degenerate.}. Rather it is closely related to the
formation of Landau levels in a uniform magnetic field. A continuum
Landau level is highly degenerate, with degeneracy equal to the number
of flux quanta piercing the plane. Thus, for the flux density here, of
one flux quantum per unit cell, a Landau level would have one state
per unit cell: that is one band within the Brillouin zone. The lowest
band of Figs.~\ref{fig:dirac}-\ref{fig:sharp}, with its narrow
width and Chern number of 1, is the optical flux lattice equivalent of
the lowest Landau level.

The above scheme can be generalized to atoms of the alkali-metal
family, whose ground state $n\,S_{1/2}$ is split in two hyperfine
levels $I\pm 1/2$, where $I$ is the nuclear spin. The laser excitation
is tuned in this case around the resonance lines $D_1$ (coupling to
$n\,P_{1/2}$ with detuning $\Delta_1$) and $D_2$ (coupling to
$n\,P_{3/2}$ with detuning $\Delta_2$).  Let us focus here on the
lowest hyperfine level $F=I-1/2$. For the configuration of
Fig.\,\ref{fig:scheme} the atom-laser coupling can be written
\begin{equation}
\hat V =  \frac{\hbar \kappa^2_{\rm tot}}{\bar \Delta}\hat\openone 
+  \hat {\bm F}\cdot {\bm B} ,
\label{eq:mmat}
\end{equation}
where $\bar \Delta^{-1}=(1/3) \Delta_1^{-1}+(2/3)\Delta_2^{-1}$, $\hat{\bm F}$ is the angular momentum operator in the ground state manifold
in units of $\hbar$, and 
\begin{equation}
B_x +iB_y = \xi\,E\kappa_0,\qquad B_z=\xi \,(|\kappa_-|^2-|\kappa_+|^2),
\label{}
\end{equation}
with $\xi=(\Delta_2^{-1}-\Delta_1^{-1})\hbar/[3(F+1)]$. Under the
unitary transformation $\hat{U} \equiv \exp(-i {\bm k}_3 \cdot {\bm
  r}\, \hat{F}_z)$ the Hamiltonian takes a similar form to
(\ref{eq:gauge}), now with $\hat{\sigma}_z/2$ replaced by $\hat{F}_z$,
and again with a coupling $\hat{V}'$ in which $\kappa_0$ is replaced
by $\kappa_0' =  e^{-i{\bm k}_3\cdot {\bm r}} \kappa_0$ giving the unit cell of the honeycomb lattice as
before.  Adiabatic motion of the atom still leads to a dressed state
with angular momentum along the vector ${\bm n}$ that wraps the Bloch
sphere once in the unit cell. However, now the Berry phase acquired is
larger by a factor of $2F$\cite{Berry:1984}. Therefore, the unit cell
contains $N_\phi = 2F$ flux quanta.  This increase of $N_\phi$ leads
to an important new feature: a continuum Landau level now corresponds
to $N_\phi= 2F$ states per unit cell.  Thus, the analogue of the
lowest Landau level is a set of $2F$ low-energy bands with a net Chern
number of 1. Spatial variations in the scalar potential and flux
density will cause these bands to split and to acquire non-zero
widths.

We shall illustrate the physics for $F>1/2$ by describing the
properties for bosonic atoms with $F=1$. This is a very important
case, as it applies for example to the hyperfine ground states of
$^{23}$Na, $^{39}$K, and $^{87}$Rb. In order to control the respective
role of the various Zeeman sublevels, we also include in the
Hamiltonian the additional Zeeman coupling to an external magnetic
field
\be \hat{W}_\gamma= \gamma_1 {\hat{F}_z} + \gamma_2
\left(1-{\hat{F}_z}^2\right) 
\ee 
where $\gamma_1$ and $\gamma_2$ correspond to the linear and quadratic
Zeeman effects, respectively.  For simplicity we assume that
$|\Delta_1|\ll |\Delta_2|$ and we take into account only the coupling
between the ground state and the $P_{1/2}$ excited state ($D_1$
line). As above, we characterize the atom-laser coupling by the angle
$\theta$ and the two dimensionless parameters $\epsilon=E/\kappa$ and
${\cal V}/\Er$, where ${\cal V}=\hbar\kappa^2/(3\Delta_1)$.

\begin{figure}
\includegraphics[width=0.98\columnwidth]{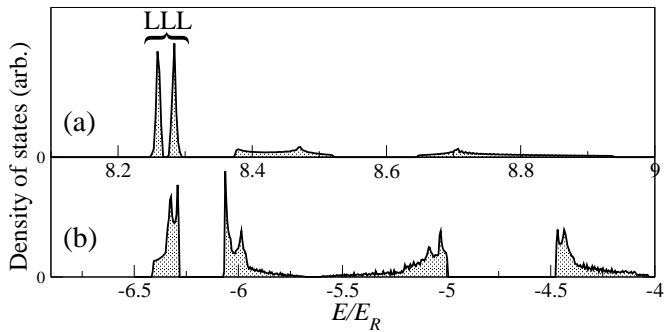}
\caption{Density of states for $F=1$. 
(a) ${\cal V}=8\,\Er $, $\epsilon=0.8$, $\theta = 0.2$,  $\gamma_1 = 0$, $\gamma_2 =  -0.15\,\Er$. The two lowest bands have
  Chern numbers of $-1$ and $2$. Taken together, these are analogous to
  the lowest Landau level (LLL).
  (b) ${\cal V}=2\,\Er $, $\epsilon=3$, $\theta = 0.5$, $\gamma_1 = 10\,\Er$, $\gamma_2 = -10\,\Er$.
The system behaves as a two-level
  system formed from $m=0,-1$.  The lowest energy band is analogous to the LLL.
  \label{fig:f1}}
\end{figure}

For a state-independent potential ($\epsilon
=\theta=\gamma_1=\gamma_2=0$) and positive detuning $\Delta_1$, each
of the three states $|F=1,m\rangle$ with $m=0,\pm 1$ experiences a honeycomb lattice potential and
has two Dirac points. Now the offsets in wavevector are such that
there are three special ${\bm k}$-points, at each of which two Dirac 
points coincide.  Again the Dirac points are split by perturbations
that break time-reversal symmetry ($\epsilon,\theta\neq 0$) and the
lowest three bands can acquire non-zero Chern number. 
Differences from $F=1/2$ emerge in the regime of intermediate
coupling.  In particular we find a domain in parameter space where the
two lowest energy bands are closely spaced, with narrow total width
and a net Chern number of 1 [see Fig.\,\ref{fig:f1}(a)]. This confirms
the expectation mentioned above that the set of the $2F=2$ lower bands
constitutes in the case the analogue to the lowest Landau level.
One can also make use of the quadratic Zeeman splitting $\gamma_2 \neq
0$ to arrange situations in which only two of the $m$-states
contribute. This two-level system is then of the same form as $F=1/2$.
In this case, parameters can be chosen to form a single narrow lowest
energy band with Chern number 1, which by itself is analogous to the
lowest Landau level.  An example is presented in
Fig.\,\ref{fig:f1}(b), showing similar features to $F=1/2$
(Fig.\,\ref{fig:dirac}).

These narrow bands with non-zero Chern number are excellent
systems with which to explore strong-correlation phenomena related to fractional
quantum Hall physics \cite{Cooper:2008}, both for fermions and
bosons.  Given the contact nature of interactions in ultracold atomic gases,
one might think that the Pauli principle would prevent interparticle
interactions between fermions within a single band.  Interactions
would indeed vanish if the atoms were to move adiabatically, since
then, at each point in space, only a single dressed state would be
relevant. However, non-adiabatic effects lead to residual interactions
between fermions.  In a semi-classical picture, one can imagine two
atoms that approach the same point in space from different
directions. Since the motion is non-adiabatic, when they reach that
point they can be in different internal states, so there is some
probability that they coincide and interact.

To understand quantitatively how non-adiabatic motion of the fermions
leads to interparticle interactions within a single band, consider a
system of spin-$1/2$ fermions ($m=\pm 1/2$) with spin-independent
contact potential $V_{m,m'}({\bm r},{\bm r}') = g_{\rm 2D}
\delta^2({\bm r}-{\bm r}')$.  The mean interaction energy for any Fock
state is
 \bea
\nonumber
 \langle \hat{H}_I\rangle 
& = & 
\frac{1}{2} g_{\rm 2D}
\int \!\!\!
\sum_{m,m'} 
 \left[
\rho_{m,m}
\rho_{m',m'} - 
|\rho_{m,m'}|^2\right] 
d^2{ r}
\\
& = & 
\frac{1}{4} g_{\rm 2D} \int 
 \rho^2({\bm r}) \left[1-P^2({\bm r})\right]
d^2{ r}
\label{eq:hpol}
 \eea
where $\rho_{m,m'}({\bm r}) \equiv \langle
c^\dag_{m}({\bm r}) c_{m'}({\bm r})\rangle$ and $c^\dagger_m(\bm r)$ creates a particle at $\bm r$ in spin  state $m$. In Eq.\,(\ref{eq:hpol}) we have introduced the local total
density $\rho$ and local polarization $P$
(in units of $\hbar/2$).  In the adiabatic limit, the local
polarization is everywhere maximal, $P=1$, so the interaction energy
vanishes. However, for non-adiabatic motion the local
polarization is reduced  and the interaction energy is non-zero.

We quantify the size of the interactions by considering a {filled}
band.  Then $ \rho_{m, m'}({\bm r}) = \sum_{{\bm k}} \left[\phi^{{\bm
      k}}_{m}({\bm r})\right]^*\phi^{{\bm k}}_{m'}({\bm r})$ where
$\phi^{{\bm k}}_m({\bm r})$ is the space- and spin-dependent
wavefunction of the Bloch state ${\bm k}$ in this band.  We define an
interaction parameter $I \equiv {\langle H_I\rangle}/E_0$ with
$E_0\equiv (1/4) g_{\rm 2D} \int \rho^2\, d^2{ r}$ the mean energy for
$P=0$.  We have computed the total interaction energy for the lowest
energy band in Fig.\,\ref{fig:dirac} for each of the three
cases\footnote{Although in the first two cases there is an energy
  overlap with the second band, one can still compute the interaction
  energy for the Fock state in which the lowest energy band is
  filled.} of $(\epsilon,\theta)$. For $\epsilon=\theta=0$ the two
spin states are decoupled, so the lowest band is unpolarized ($P=0$)
and $I=1$. For weak coupling $\epsilon=\theta=0.1$ we find $I=0.95$,
showing a (small) suppression of interactions.  With increasing
coupling, the interactions continue to be suppressed.  However, even
for intermediate coupling ($\epsilon=0.4$, $\theta=0.3$), where the
lowest band has become narrow, the interaction parameter is $I=0.42$,
showing that the motion is still far from adiabatic. The ultra-narrow
band shown in Fig.\,\ref{fig:sharp} has a small value for the
interaction parameter $I=0.046$.

Since the low energy bands described above are analogous to the lowest
Landau level, leading candidates for strongly correlated phases
include the Laughlin states, at mean particle densities of
$\bar{\rho}=(1/3) N_\phi/A_{\rm cell}$ for fermions and
$\bar{\rho}=(1/2) N_\phi/A_{\rm cell}$ for bosons, with $N_\phi=2F$;
or, if the Zeeman effect is used to isolate an effective two-level
system, with $N_\phi =1$.  Strongly correlated phases of this kind are
energetically competitive when the mean interaction energy is on the
scale of the bandwidth $\Delta E\sim 0.1\,\Er $.  For bosons, taking
$E_{\rm i}=g_{\rm 2D}\int \rho^2\,d^2r \approx (1/4)g_{\rm
  2D}N_\phi^2/A_{\rm cell}$ as a typical scale for the interaction
energy, we obtain using the cell area given above $E_{\rm i}/\Delta
E\approx (1/4)N_\phi^2 \tilde g$, where $\tilde g=Mg_{\rm 2D}/\hbar^2$ is
the dimensionless interaction strength in 2D. With cold atomic gases
confined in a strong one-dimensional optical lattice along the $z$
direction, values of $\tilde g$ can approach unity
\cite{Hadzibabic:2009}, so that reaching the strongly correlated
regime can be realistically envisaged. For spin 1/2 fermions, the
partial suppression of $s$-wave interactions discussed above somewhat
complicates the realization of the correlated regime. However one can
rely on Feshbach resonances to produce a strongly interacting 2D Fermi
gas \cite{Frohlich:2011}, and thus also in this case reach the point
where $\langle H_I\rangle \sim \Delta E$. Note that in the case of
$^{171}$Yb or $^{199}$Hg, one has to turn to optically induced
Feshbach resonances \cite{Enomoto:2008,Reichenbach:2009}, because the
spin of the ground state has a purely nuclear origin hence `standard' magnetic Feshbach resonances do not occur.

The appearance of fractional quantum Hall states is predicted to have
to various consequences for the ground state and excitation spectrum of
the cold atomic gas, as discussed in the
literature\cite{Bloch:2008,Fetter:2009,Cooper:2008}.  Perhaps the most
evident feature is a plateau (or plateaus) in {\it in situ} images of
the density of a trapped gas, arising from the incompressibility of
the FQH states. For the Laughlin states, these plateaus are expected
at mean densities for which there is no competing Mott insulator
state. 

In summary we have presented a simple scheme by which optical flux
lattices can be implemented using two-photon dressed states. The
method applies to commonly studied atomic species, requires few
lasers, and uses only well-established techniques. Our studies show
that for practical parameters, the non-adiabatic character of atomic
motion can lead to significant interactions even for fermions
occupying a single band.  This scheme should allow experiments on cold
atomic gases to explore strong correlation phenomena in regimes of
flux density previously unattainable.

\acknowledgments{This work was supported by EPSRC Grant EP/F032773/1, IFRAF, ANR (BOFL project), and E.U. (MIDAS project). J.D. acknowledges useful discussions with F. Gerbier and N. Goldman.}

\end{document}